\documentclass[12pt,preprint]{aastex}

\slugcomment{Submitted to ApJ }

\shorttitle{Fe K$_{\alpha}$ line study in Non-magnetic CVs}

\shortauthors{Rana et al.}

\begin{document}


\title{Study of Fe K${\alpha}$ lines in Non-magnetic Cataclysmic
Variables using \emph{Chandra} HEG data}


\author{V.R. Rana\altaffilmark{1} and K.P. Singh}
\affil{Department of Astronomy and Astrophysics, Tata Institute of
Fundamental Research, Homi Bhabha Road, Mumbai - 400 005, INDIA}
\email{vrana@tifr.res.in, singh@tifr.res.in}

\author{E.M. Schlegel\altaffilmark{2}}
\affil{Harvard-Smithsonian Center for Astrophysics, 60 Garden Street,
Cambridge, MA 02138, U.S.A.}
\email{eschlegel@cfa.harvard.edu}

\and

\author{P.E. Barrett\altaffilmark{3}}
\affil{Space Telescope Science Institute, ESS/Science Software Group,
Baltimore, MD 21218, U.S.A.}
\email{barrett@stsci.edu}


\altaffiltext{1}{Joint Astronomy Programme, Department of Physics,
Indian Institute of Science, Bangalore 560~012, INDIA}
\altaffiltext{2}{Current address: Department of Physics and Astronomy,
University of Texas-San Antonio, 6900 North loop 1604 west, San
Antonio, TX, 78249, USA \\
Email: eric.schlegel@utsa.edu}
\altaffiltext{3}{Current address: Department of Physics and Astronomy,
The Johns Hopkins University, Baltimore, MD 21218, USA \\
Email: barrett@pha.jhu.edu}


\begin{abstract}
Results from a study of Fe K${\alpha}$ emission lines for a sample of
six non-magnetic Cataclysmic Variables (CVs) using the high resolution
X-ray data from the \emph{Chandra} High Energy Transmission Grating
(HETG) are presented.  Two of the sources, SS~Cyg and U~Gem are
observed in both quiescent and outburst states whereas V603~Aql,
V426~Oph, WX~Hyi and SU~UMa are observed only in quiescence. The
fluorescent Fe line is prominent in V603~Aql, V426~Oph and SS~Cyg
during quiescence indicating the presence of a conspicuous reflection
component in these sources.  The observed equivalent width of the
fluorescent Fe line is consistent with reflection from a white dwarf
surface that subtends 2$\pi$ solid angle at the X-ray source. During
the outburst in SS~Cyg, the fluorescent line is red-shifted by about 2300
km s$^{-1}$.  The Fe XXV triplet at 6.7 keV is found to be dominant in
all sources. The value of the G-ratio derived from the Fe XXV triplet 
indicates that the plasma
is in collisional ionization equilibrium during the quiescent state.
The Fe XXV line is significantly broadened in U~Gem and SS~Cyg during the
outbursts compared to quiescence, indicating the presence of a high
velocity material near the white dwarf during the outburst.  The ratio
of Fe XXVI/XXV indicates a higher ionization temperature during
quiescence than in outburst in U~Gem and SS~Cyg.
\end{abstract}





\section{Introduction}

Non-magnetic cataclysmic variables (CVs) are a subclass of CVs in
which a white dwarf with a weak magnetic field (B$\lesssim$10$^{4}$G)
accretes material from the Roche lobe of a late type dwarf companion
star \citep{war95}. The accretion takes place via a disk around the
white dwarf.  Due to the weakness of the magnetic field the disk
extends to the surface of the white dwarf. The Keplerian velocity of
the material in the disk is generally greater than the rotational
speed of the white dwarf.  Near the inner edge of the disk the
material slows down to match the white dwarf rotation.  The X-ray
emission is thought to arise from this boundary layer between the
white dwarf and the inner edge of the accretion disk.

Dwarf Novae (DNe) are a subclass of the non-magnetic CVs that show
frequent (at the interval of weeks to months) outbursts.  According to
theory, a thermal--viscous instability in the disk produces repetitive
outbursts in these systems \cite[see][for a review]{lasota01}.  On the
other hand, in classical novae, another subclass of non-magnetic CVs,
the eruptions are due to thermonuclear runaway of hydrogen rich
material accreted on to the white dwarf. The amplitude of outburst in
these systems is greater than in DNe.

A sample of DNe were studied using low-resolution (E/$\bigtriangleup$E 
$\simeq$ 5 @ 6 keV) \emph{EXOSAT} medium
energy (ME) data \citep{mukai93} and it was found that line emission
near 6.7 keV is a common feature in the hard X-ray spectra of DNe.
The Fe XXV triplet is one of the most intense set of lines in the
hard X-ray spectra of CVs \citep{pandel05}.  The line emission near
6.7 keV originates from Fe K${\alpha}$ emission that has three main
components: a fluorescent line at 6.41 keV, Fe XXV line (He-like) that
has four subcomponents (a resonance line `$r$' at 6.7002 keV, two
intercombination lines `$i_{1}$' and `$i_{2}$' at 6.6821 and 6.6673
keV, respectively and a forbidden line `$f$' at 6.6364 keV), and Fe
XXVI Ly$_{\alpha}$ line (H-like) with two subcomponents at 6.973 and
6.952 keV.  The Fe XXV triplet ($r$, $i$, and $f$) is due to the
transitions 1s2p $^{1}$P$_{1}$ $\longrightarrow$ 1s$^{2}$
$^{1}$S$_{0}$, 1s2p $^{3}$P$_{2,1}$ $\longrightarrow$ 1s$^{2}$
$^{1}$S$_{0}$, and 1s2s $^{3}$S$_{1}$ $\longrightarrow$ 1s$^{2}$
$^{1}$S$_{0}$.  The two components of the Fe XXVI line are due to the
transitions 2p $^{2}$P$_{3/2}$ $\longrightarrow$ 1s $^{2}$S$_{1/2}$
and 2p $^{2}$P$_{1/2}$ $\longrightarrow$ 1s $^{2}$S$_{1/2}$.  The Fe
XXV and XXVI lines come from a plasma having a temperature of
10$^{7-8}$ K. The fluorescent line originates from relatively cold
iron (Fe I--XVII) having temperatures $\leq$10$^{6}$ K.  The
low-resolution of the \emph{EXOSAT} ME data was insufficient to
distinguish between the various components of Fe K$_{\alpha}$
emission.  Moderate resolution (E/$\bigtriangleup$E = 50 @ 6 keV) 
spectroscopy with \emph{ASCA} and
\emph{XMM-Newton} has been a little more successful in resolving the
components of Fe K$_{\alpha}$ emission, and several authors have
reported Fe K${\alpha}$ lines for individual non-magnetic systems
using the \emph{ASCA} \citep[see][for SS~Cyg]{done97} and
\emph{XMM-Newton} data \citep[see][for YZ~Cnc and references
therein]{hakala04}. Recently, \cite{baskill05} presented a study of 34
non-magnetic CVs using \emph{ASCA} data.  The CCD spectrometer (SIS)
allowed them to separate the three Fe K$_{\alpha}$ lines in the
spectra.  Of the 34 objects, only 4 bright sources showed the presence of a
prominent fluorescent Fe line at 6.4 keV.  This feature is attributed
to the reflection of X-rays from the white dwarf surface and/or the
accretion disk.

High spectral resolution (E/$\bigtriangleup$E $>$200 @ 6 keV) data 
obtained with \emph{Chandra} provides an
opportunity to study in detail the strength and profile of these
lines. The Fe XXV and XXVI emission lines that originate from
the high temperature (10$^{7-8}$ K) plasma serve as an important
diagnostic tool for temperatures near the shocked regions in the
boundary layer of these CVs.  The intensity ratio of Fe XXVI to Fe
XXV lines provides a measure of the ionization temperature of the
plasma.  Intensity ratios defined using the Fe XXV triplet can be used
in principle 
to get information about the ionization state, temperature and density
of the emitting plasma in collisional ionization equilibrium.
Specifically, the ratio G, defined as ($f$+$i$)/$r$ is sensitive to
the electron temperature as well as the ionization state of the
plasma.  The ratio R=$f$/$i$ provides a density diagnostic.  The
fluorescent line of Fe that comes from relatively colder material can
provide information about the contribution of reflected hard X-rays
from the illuminating region.

In this paper, we present a detailed study of Fe K${\alpha}$ emission
lines from a sample of non-magnetic CVs using the best available
energy resolution X-ray data available from \emph{Chandra} grating
instruments.  The next section describes the observations and the data
analysis procedure.  The results are presented in \S 3, and are
discussed in \S 4.  The conclusions are summarized in \S 5.

\section{Observations and Data Analysis}

High resolution X-ray data for a sample of six non-magnetic CVs
obtained from the \emph{Chandra} \citep{weiss02} High Energy
Transmission Grating \cite[HETG;][]{markert94} during 2000 August to
2002 December are analyzed.  The observations were made with HETG in
combination with Advanced CCD Imaging Spectrometer
\cite[ACIS;][]{garmire03} in faint spectroscopy data mode.  The sample
consists of five dwarf novae (SS~Cyg, U~Gem, WX~Hyi, SU~UMa and
V426~Oph) and one old nova (V603~Aql).  A log of the observations for
the sources studied is given in Table 1.  Average spectra for five of
these sources in quiescence have been presented previously
\citetext{for SS~Cyg, U~Gem and V603~Aql see
\citealp{mukai03,mauche05,mukai05,szkody02}, for V426~Oph see
\citealp{homer04}, and for WX~Hyi see \citealp{perna03}}.
Two of the six sources, SS~Cyg and U~Gem, are also observed during
outbursts.  SS~Cyg was observed twice during a short outburst in the
optical: near the peak of the outburst on 2000 September 12 and during
an early decline of a narrow outburst on 2000 September 14
\citep{mauche05}.  U~Gem was observed at the peak of an outburst 
\citep{mauche05}. SU~UMa was observed twice during 2002 April 12 \&
13.  The individual spectra of SU~UMa show identical features, we have
therefore combined the two spectra.  A total of ten observations are
analyzed.

\begin{figure}
\includegraphics[angle=-90,scale=.60]{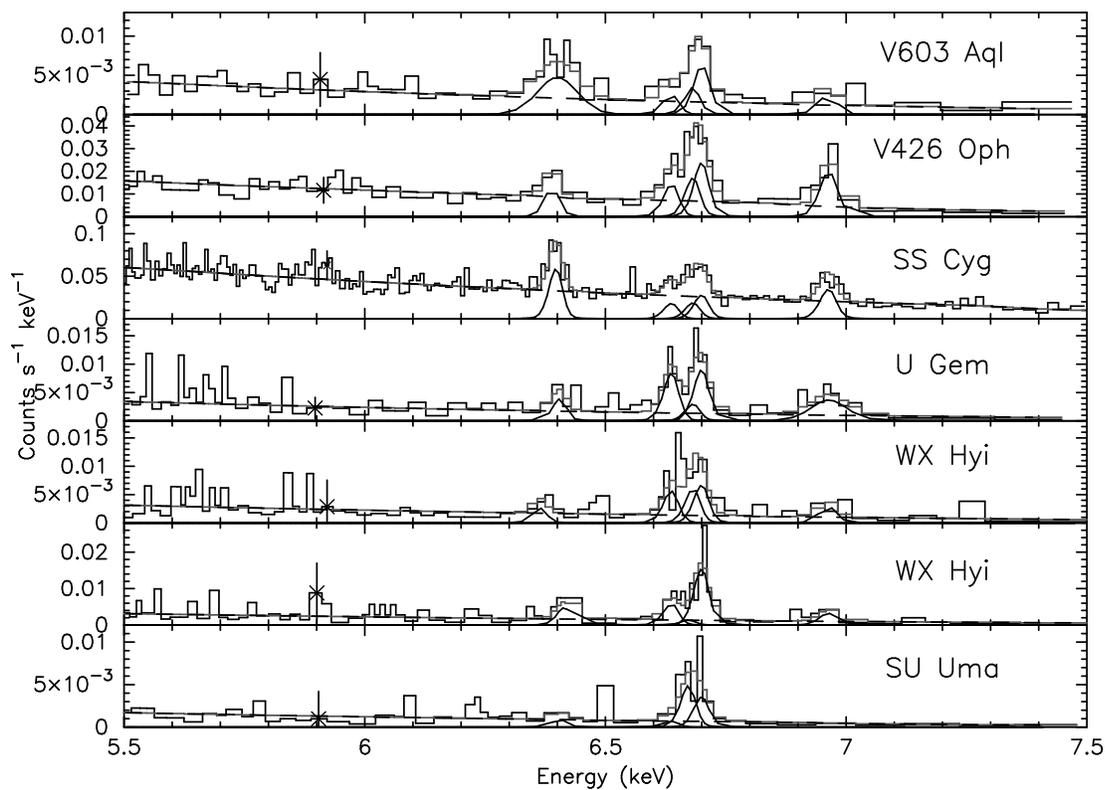}
\caption{
\emph{Chandra} HEG spectra of six non-magnetic
CVs in quiescence.  The fitted line components are the fluorescent
(6.41 keV), He-like (forbidden, intercombination, and resonance lines
at 6.6364, 6.6750, 6.70 keV), and H-like (6.96 keV) emission lines of Fe.
A single Gaussian is used for the unresolved He-like intercombination
and H-like doublets.
\label{fig1}}
\end{figure}

The data are reduced using the \emph{Chandra} Interactive Analysis of
Observations (CIAO version 3.1) software package following the
analysis threads for HETG/ACIS observations \footnote[1]{(see
\url{http://cxc.harvard.edu/ciao/threads/gspec.html})}.  For each
source we extract the HEG (high energy grating) and MEG (medium energy
grating) spectra for the combined $\pm$1 orders to improve the
statistics of the spectra.  The response files for each of the sources
are generated using the \textit{fullgarf} script.  The
\textit{add\_grating\_orders} script is used to sum the grating
orders and the corresponding response files. The HEG has an
energy resolution of 43 eV or $\sim$1900 km s$^{-1}$ (FWHM) at 6.7 keV
energy.  The HEG has a higher effective area than the MEG in the 6--7
keV band, so we have used the HEG data to study the iron lines.  We
show in Figures \ref{fig1} \& \ref{fig2} only the relevant portion
(5.5--7.5 keV) of the HEG spectrum of each source, since our focus is
the study of the Fe XXV, XXVI, and fluorescent emission lines.
Representative error bars for each spectrum are shown at a point near
5.9 keV.  The default binning for the HEG spectra is 0.0025 $\AA$,
which oversamples the instrumental resolution by a factor of $\sim$5.
For the line profile modeling using grating response, oversampling is
recommended.  Therefore, we have tried several ways of binning the
data (for example binning by a factor of 2, 5, and 10 and minimum 5 or
10 counts per bin) and found that the binning by a minimum of 5 or 10
counts per bin preserves the instrumental resolution and line profile
at that energy.  This binning groups the energy channels corresponding
to the continuum by a large factor but the channels for lines remain
single or are grouped by a factor of two only.  Hence, the spectra
were binned to give a minimum of 5 or 10 counts per bin depending on
the source strength.

For simplicity, a power law is used to model the local continuum while
the emission lines are modeled using several Gaussians.  For spectral
analysis the XSPEC v11.3.1 \citep{arnaud96} and ISIS v1.2.6
\footnote[2]{\url{http://space.mit.edu/CXC/ISIS}} software packages
are used.  A Gaussian is used to fit the fluorescent Fe line.  The
$r$, $i$, and $f$ components of Fe XXV lines are fitted with three
Gaussians (see \S 3.2 below).  The Fe XXVI lines are fitted with a
single Gaussian, since the two Ly$_{\alpha}$ components are unresolved
at the resolution of the \emph{Chandra} HEG.  To check for any
broadening of the emission lines, other than that of the instrument,
the width of the Gaussian is allowed to vary during the initial
fitting.  For final fitting the width is fixed at zero for the cases
where it is found consistent with the instrumental resolution.  The
possible shift in the line position is also checked by allowing the
line energy to vary.  The line centroids are kept fixed wherever they
are found to be consistent with the expected values of line energy.
The measured equivalent width (EW) and the photon flux in the lines
for all sources are listed in Table \ref{tbl-2}.  The errors are with
90\% confidence interval for a single parameter
($\Delta\chi^{2}$=2.706).  Since the counts in the HEG spectra are
low, we have used the C statistic \citep{nousek89} based on the
maximum likelihood approach, which provides a better defined
confidence range for a parameter derived using data with a small
number of counts per energy bin.

\section{Results}
The results of our analysis are summarized in Tables \ref{tbl-2} \&
\ref{tbl-3} and in Figures \ref{fig1} \& \ref{fig2}.  The details are
presented in the following subsections.

\subsection{The Fluorescent Iron Line}
During the quiescent state, the fluorescent Fe line at 6.4 keV is
prominent in V603~Aql, V426~Oph, and SS~Cyg, whereas it is relatively
weak in U~Gem, WX Hyi, and SU~UMa, as shown in Fig. \ref{fig1}.  The
width of the fluorescent Fe line is consistent with the instrumental
resolution in all sources, except in V603~Aql, which shows the
strongest fluorescent Fe line with an EW of 162$^{+99}_{-65}$ eV.  The
broadening in the line is a factor of $\sim$2.5 greater than the instrumental
resolution.  For WX~Hyi during 2002 July 28 observation the best
fit value for the fluorescent Fe line energy is 6.36$\pm$0.04 keV.
Though the best fit value indicates an apparent red-shift in the line,
the upper value of the 90\% confidence interval is consistent with a
zero redshift.
The other observation of WX~Hyi during 2002 July 25 shows the line
center at its expected value of 6.4 keV with an EW of 55$^{+44}_{-30}$
eV.  For the other sources the line center is at its expected value of
6.4 keV during the quiescent state.  In V426~Oph, SS~Cyg, and U~Gem,
the fluorescent Fe lines with EWs of 39$^{+22}_{-18}$,
59$^{+25}_{-20}$ and 46$^{+31}_{-24}$ eV, respectively, are identical
within the 90\% confidence limit.  The EW of the line in SU~UMa has a
90\% confidence interval of 0--70 eV with the best fit value of 23 eV,
and is the weakest measured among the six sources in the quiescence.

The principal components of the Fe K$_{\alpha}$ emission lines during
the U~Gem and SS~Cyg outbursts are shown in Figure \ref{fig2}.  U~Gem
shows a weak fluorescent line with an EW of 22$^{+16}_{-14}$ eV
centered at 6.4 keV. On the other hand, SS~Cyg shows a red-shifted
fluorescent line from Fe during the two observations. The best fit
value of the line center during 2000 September 12 is 6.35$\pm$0.02 keV
and during 2000 September 14 it is 6.35$\pm$0.01 keV with their
equivalent widths of 68$^{+26}_{-40}$ and 60$^{+14}_{-14}$ eV,
respectively. The shift in the line energies correspond to
velocities of 2300$^{+980}_{-900}$ and 2300$^{+500}_{-440}$ km
s$^{-1}$ during 2000 September 12 and 14 observations for SS~Cyg,
respectively. The width of the line is consistent with the
instrumental resolution in both sources during the outburst.  The EWs
are about the same in outburst and quiescence.

\subsection{Fe XXV Line}
The Fe XXV line is a dominant line in all sources during both the
quiescent and outburst states.  At the \emph{Chandra} HEG resolution,
the $r$ and $f$ components are clearly resolved.  The separation
between $i_{1}$ and $i_{2}$ components is only 14.8 eV and these lines are
unresolved, hence a single Gaussian with center at 6.675 keV has been
used to account for these components.  The $i$ component is marginally
resolved from $f$.  Thus, we used three Gaussians to characterize the $r$, 
$i$, and $f$ components of Fe XXV line. 
It is possible to discern some of the
structure in the line components and get information about their
relative strengths and probe the density
and temperature structure of the emitting region.

We have checked for any shift in the line positions and line
broadening for these components and found that line centroids are
consistent with the expected values and line width with the
instrumental resolution for all sources in quiescence.  Among the
three components of the Fe XXV lines, the resonance line is the
stronger one in all sources in the quiescent state, except SU~UMa.

\begin{figure}
\includegraphics[angle=-90,scale=.60]{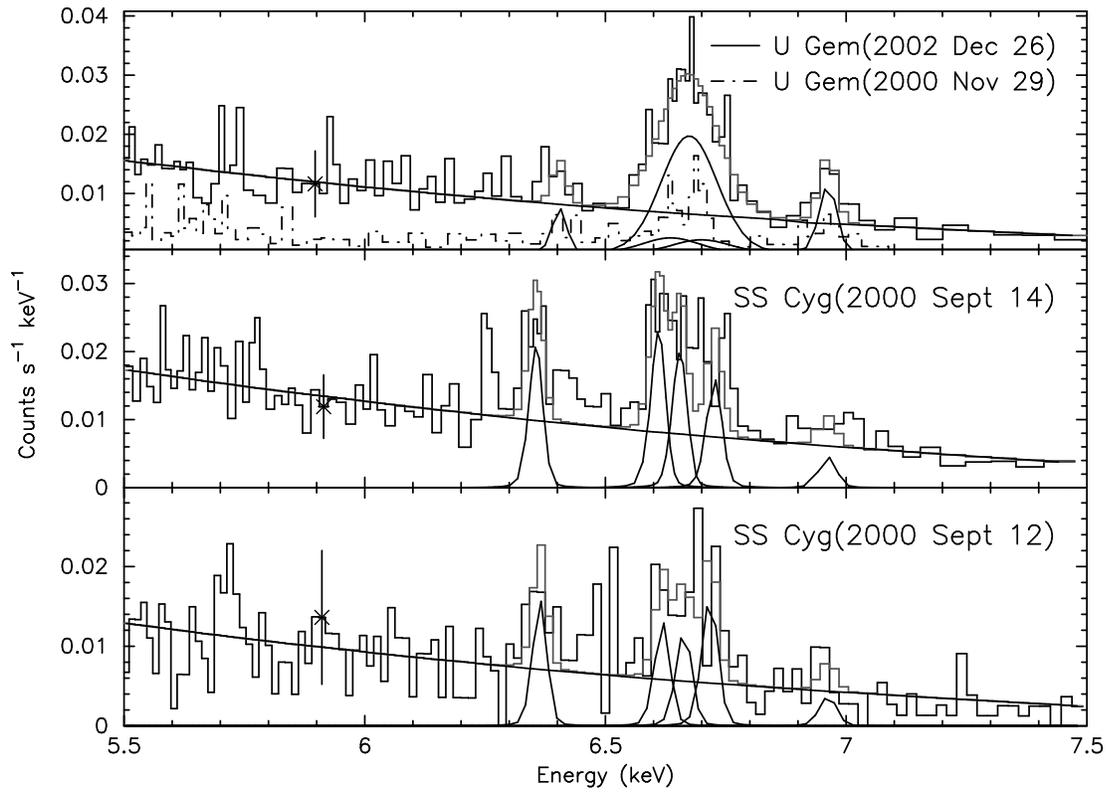}
\caption{Same as Figure \ref{fig1} but during the outbursts of U~Gem
and SS~Cyg. The dash-dotted curve in the top panel for U~Gem
represents the spectrum in quiescence taken from Fig. \ref{fig1} 
for comparison.
\label{fig2}}
\end{figure}

During outbursts, the systems show considerably different profiles for
the Fe XXV lines.  An inspection of this line indicates that the $r$,
$i$ and $f$ components are either broadened or shifted significantly
from their expected values (see Fig. \ref{fig2}). The broad
emission lines in the outburst spectra of SS~Cyg and U~Gem have been
previously reported by \cite{mauche05}. For comparison we show both
the outburst (solid line) and the quiescence (dash-dotted line)
spectra of U~Gem in the top panel of Fig. \ref{fig2}.  In order to
quantify any broadening or shift, we have first fitted the Fe XXV line
components with unresolved Gaussians and centroids at the
theoretical positions.  This gives significant residuals near the wings
of the line for both sources indicating that the $r$, $i$ and $f$ line
components are either broadened or shifted.  Hence some modification
must be made to the base model of ``no broadening and no shift'' to
reproduce the data for these two sources.

After fitting the lines of U~Gem with a Gaussian of zero width,
significant residuals are seen near 6.6 and 6.75 keV indicating the
presence of broad wings in the Fe XXV line.  We then allowed the widths
of the $r$, $i$ and $f$ lines to vary in order to account for the
residuals.  However, since the Fe~XXV triplets arise from transitions in a 
single ion, they must be physically related, and thus likely to 
originate from
the same physical conditions of temperature and density. 
Therefore, the line widths were tied together and allowed to vary 
by an equal amount during the fitting.  The centroid of all the components
were fixed at their expected values.  The best fit with various
Gaussian components is shown in the top panel of Fig. \ref{fig2}.
This gives a common width $\sigma$=55$^{+13}_{-19}$ eV for the $r$, $i$
and $f$ components with 90\% confidence. The EW of $r$, $i$ and $f$ lines 
are 26$^{+48}_{-26}$, 325$^{+117}_{-325}$ and 28$^{+56}_{-28}$ eV, 
respectively (see Table~\ref{tbl-2}).     
Alternatively, it is also possible to fit the Fe XXV lines by allowing
the center energy of $r$, $i$ and $f$ components to vary, but retaining
the unresolved lines.  However, the former fitting leads to a lower
$\chi^{2}$ by $\sim$12 and better represents the line profile.

Of the two outburst observations of SS~Cyg, the one on 2000 September
14 has a longer exposure and better statistics than on 2000 September
12.  The September 14 observation shows a flat top profile for the Fe
XXV lines unlike the broad Gaussian shape seen in U~Gem during
outburst (see Fig. \ref{fig2}, middle panel).  We provided the
same treatment described above for U~Gem to the SS~Cyg spectra.
Fixing the line energies at their expected values and
allowing the width of the $r$, $i$ and $f$ lines to vary, fails to account
for the flat portion of the line that spans $\sim$6.6--6.8 keV range.
This suggests that the
line positions are shifted rather than being broadened.  Therefore, we
allow the central energies of the $r$, $i$ and $f$ lines to vary, 
retaining the unresolved line widths.
The $r$, $i$ and $f$ lines are allowed to shift in the same direction
and by an equal amount. 
However, this fit gives significant residuals at either the blue or the 
red side of the overall profile and fails to account for the flat top
shape of the line.  A better fit can be obtained by allowing the centroid
of the individual lines to vary in either direction by equal amounts.
This fit, shown in the middle panel of Fig.~\ref{fig2}, leads to an unusual 
situation in which the $r$ line shows a blue-shift of 25 eV
(1120$\pm$224 km s$^{-1}$), and the 
$i$ and $f$ lines are red-shifted by the same amount.  The best fit values
of energies for $r$, $i$ and $f$ lines are respectively
6.725, 6.650, and 6.612 keV with a 90\% confidence interval of $\pm$0.005
keV.  The equivalent widths are
70$^{+25}_{-10}$, 72$^{+19}_{-17}$, and 82$^{+19}_{-19}$ eV for $r$,
$i$, and $f$, respectively.  These results are discussed in section 4.4.

Similarly, during 2000 September 12, SS~Cyg shows a flat top line
profile and hence the broadened line components do not reproduce the
observed line profile.  In this case too, a better fit to the overall 
line structure (see Fig~\ref{fig2}, bottom panel) is obtained with a  
blue-shifted $r$ line and red-shifted $i$ and $f$ lines!  
The best fit values of energies for $r$, $i$ and $f$ lines are
6.718, 6.657, and 6.619 keV, respectively, with a 90\% confidence interval
of $\pm$0.005 keV, and a shift of $\sim$18 eV (806$\pm$224 km s$^{-1}$). 
The equivalent widths for $r$,
$i$, and $f$ are 99$^{+54}_{-36}$, 60$^{+34}_{-34}$, and
65$^{+40}_{-28}$ eV, respectively.

Since the Fe XXV triplet provides important spectral diagnostics of
the emitting region, we have calculated the line intensity ratios G
and R for all sources during both the quiescent and outburst states,
and listed them in Table \ref{tbl-3}.

\subsection{Fe XXVI Line}
During the quiescent state, the center of Fe XXVI line is found to be
at 6.96 keV for all sources, indicating no shift in the line position
with respect to the laboratory value. The line width is also
consistent with the instrumental resolution except for U~Gem for which
the line is found to be broadened by a factor of $\sim$2 compared to
the instrumental value.  This indicates the presence of randomized
motion of material in the boundary layer of this source. The EW
of Fe XXVI line in U~Gem is 181$^{+71}_{-58}$ eV. It should be noted
that the dielectronic recombination produces satellite lines red-ward of
the principal lines (6.91--6.95 keV), 
which can shift the centroid toward lower energies. 

V426~Oph has a strong Fe XXVI line with an EW of 136$^{+47}_{-39}$ eV
indicating the presence of high temperature plasma in the boundary
layer.  On the other hand, this line is absent in SU~UMa, hence the
temperature of the emitting region is relatively low in this system as
compared to the other sources.  The other sources show a relatively weaker
Fe XXVI line with EW$<$100 eV.

During outburst, the Fe XXVI line is found to be at its expected value
in SS~Cyg and U~Gem.  It is also narrow and consistent with the
instrumental resolution, thus requiring no extra broadening.  This
line is stronger in U~Gem with EW of 65$^{+26}_{-23}$ eV, as compared
to SS~Cyg, which has EW of 27$^{+12}_{-12}$ and 27$^{+37}_{-24}$ eV
during the two observations.

\section{Discussion}
\subsection{Fe XXV triplet: Plasma Diagnostics}
The intensity ratios of the emission lines from He-like ion provide
valuable spectral diagnostics for temperature, density and ionization
state for plasma in collisional ionization equilibrium.  For Fe XXV
the principal lines can be contaminated by the dielectronic satellites
(DES) for plasma in collisional ionization equilibrium
\citep{oelgoetz01}.  It is not possible to resolve these lines from
the principal lines at the available energy resolution but it is
possible to infer their relative contribution to the observed spectrum
depending on the temperature of the emitting plasma.  The DES dominate
principal lines at temperatures below 3$\times$10$^7$ K and their
contribution at higher temperatures is negligible \citep{oelgoetz01}.

It can be seen from the Fig. \ref{fig1} and Table \ref{tbl-2} that the
resonance line is stronger compared to the other two components
of Fe XXV line for all sources during quiescence, except for SU~UMa.
The relative strength of the resonance line indicates that the temperature of
the emitting plasma is above 3 $\times$ 10$^{7}$ K \citep{oelgoetz01},
where the principal lines dominate.  Therefore, we can use these lines
to infer the temperature and density of the emitting region.  The
G-ratio is very close to 1 (within 90\% confidence limit; see Table
\ref{tbl-3}) for most of the sources indicating that the plasma is mainly
in collisional ionization equilibrium \citep{oelgoetz01} with electron
temperature T$_{e}$ $\geq$ 10$^{7}$ K for Fe XXV.  SS~Cyg, during one of 
the outbursts, shows a somewhat higher G-ratio value ($\sim$2.4) that might
indicate the presence of hybrid plasma in the system. For U~Gem 
during the outburst, the value of the G-ratio is unconstrained, and therefore 
does not allow us to comment on the ionization state and the temperature 
of the emitting plasma.

The mean value of the R-ratio for SS~Cyg (during both quiescence 
and outburst),
V603~Aql, V426~Oph, and WX~Hyi (2002 July 28 observation) is close to 
unity and varies between 0 to 2.5. For two sources, U~Gem and WX~Hyi 
(2002 July 25 observation) during quiescence, the R-ratio is very high 
and essentially
unconstrained (see Table~\ref{tbl-3}).  On the other hand SU~UMa and
U~Gem (during outburst) show very low values of the R-ratio.  
According to the theoretical curves presented by \cite{bau00}, the 
Fe~XXV R line ratio is $\sim$1 for low densities and rolls over to 0 
at a critical density of $\sim$1$\times 10^{17}$ cm$^{-3}$.  Thus, 
the above mentioned values of the R-ratio (also see Table~\ref{tbl-3}) for 
the Fe~XXV triplets 
do not allow us to constrain the plasma 
densities in the non-magnetic CVs studied here. 
It should be noted, however, that the Fe~XXV triplets are not fully resolved
with the HEG data and hence the error estimations on individual line
fluxes are not completely independent of each other. This may affect
the errors on the G and R-ratios.  Better resolution data with
sufficient signal to noise ratio are required to properly constrain the line 
ratios for these sources. 
  
\subsection{Fe XXVI to XXV line ratio}
In Figure~\ref{fig4}, we show the observed line ratio of Fe
XXVI/Fe XXV in various sources as a function of ionization
temperature.  We have used the summed flux of $r$, $i$ and $f$
lines of Fe XXV for calculating this ratio.  The solid curve
represents the expected line ratio for Fe as a function of ionization
temperature assuming the plasma is in collisional ionization
equilibrium \citep{mewe85}.

It is clear from the Fig.~\ref{fig4} that the lower limit of the
ionization temperature is not constrained by the HEG data for several
sources, namely SU~UMa, V603~Aql, WX~Hyi (2002 July 28), 
and SS~Cyg during one of the outburst observations.  
However, it is well constrained
for U~Gem, SS~Cyg during quiescence, and V426~Oph.  In general the
ionization temperature for all CVs studied here is $\lesssim$12 keV.
For SS~Cyg, it has been observed that during optical outburst the hard
X-ray (3--20 keV) bremsstrahlung temperature is lower by a factor of
$\sim$2 than during quiescence \citep{done97,wheatley03,mcgowan04}.
Since the plasma lines of Fe originate from the hot shocked region,
the ionization temperature derived from these lines is expected to
show a similar behavior.  The observed decrease in the ionization
temperature during outburst is consistent with previous hard X-ray
observations of the continuum in SS~Cyg.

\subsection{Reflection and the Fluorescent Fe line}
In non-magnetic CVs, the fluorescent iron line is believed to arise
due to reflection of hard X-rays from the white dwarf surface or the
inner edge of the accretion disk.  In the quiescent state at low
accretion rates, the inner accretion disk is either absent or
optically thin and hence contributes little to the observed reflection
component.  Therefore, a significant contribution to the fluorescent
Fe line in DNe in the quiescent state comes from the reflection off
the white dwarf surface. It can be seen from the Table \ref{tbl-2}
that the fluorescent line in all sources is not very strong (with
EW$\leq$60 eV) except for V603~Aql.  This could be because only the
surface of the white dwarf contributes to the fluorescent line.

\begin{figure}
\includegraphics[angle=-90,scale=0.8]{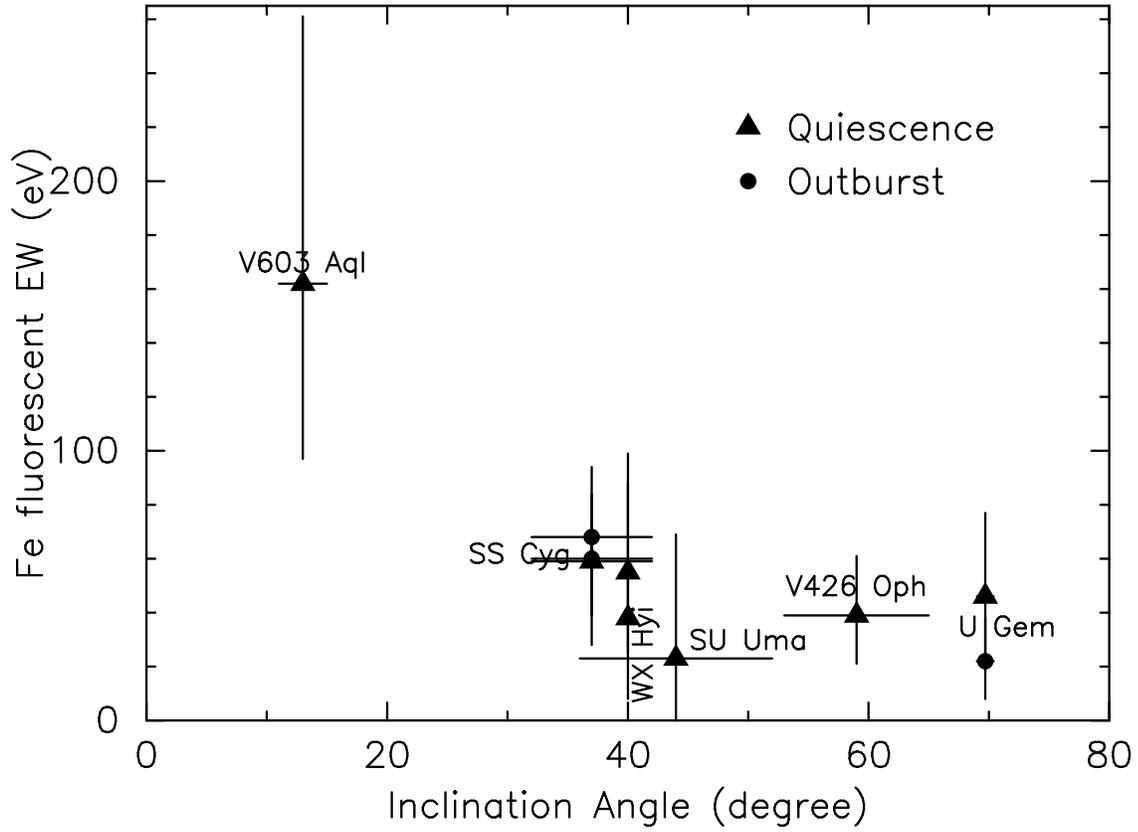}
\caption{Equivalent widths of the fluorescent Fe line as a function of
the inclination angle of binaries during quiescence and
outbursts. \label{fig3}}
\end{figure}

\begin{figure}
\includegraphics[angle=-90,scale=0.8]{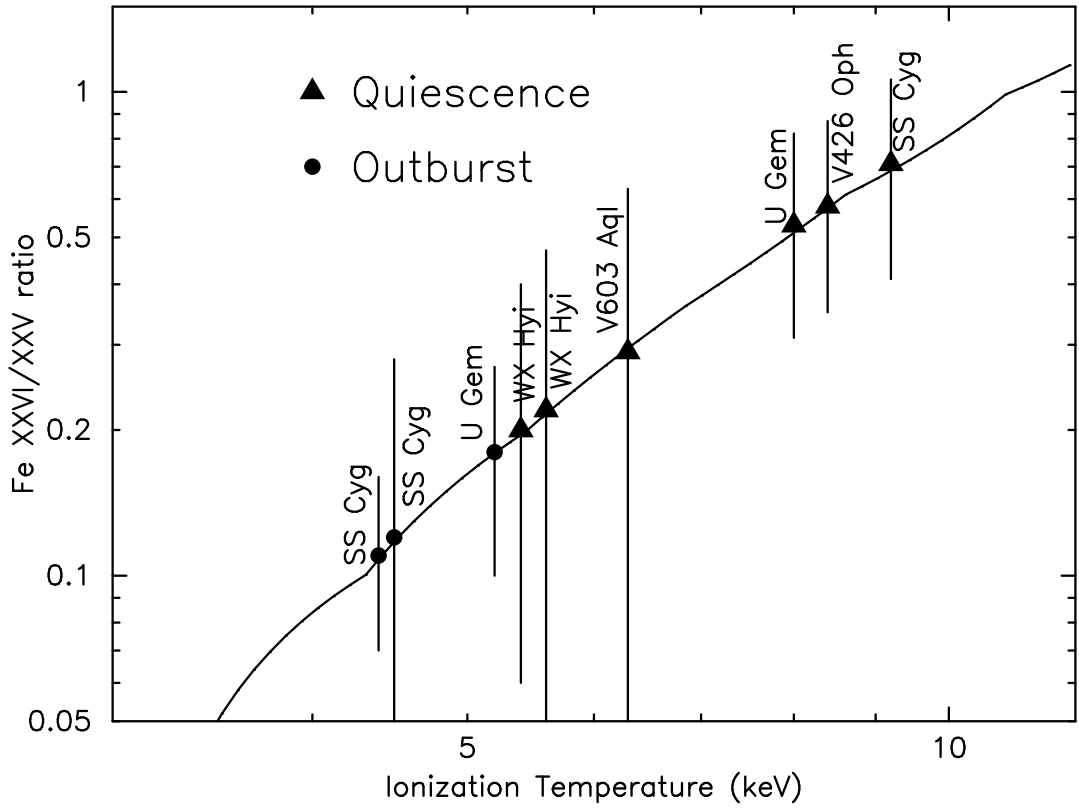}
\caption{Fe XXVI/Fe XXV line ratio as a function of ionization
temperature assuming plasma in collisional ionization
equilibrium. \label{fig4}}
\end{figure}

If the fluorescent line is mainly due to reflection, then the strength
of this line should depend on the inclination angle of the system.
For a system having a large inclination angle the fluorescent line
is expected to be weak, because it prevents the observation of any
reflection from the inner edge of the disk \citep{ramsay01}.  We have
looked for such correlations in this sample of CVs.  In Figure
\ref{fig3} we have plotted the observed EW of fluorescent line as a
function of system inclination.  The values of the inclination angle
have been taken from \cite{baskill05} and \cite{war95}.  As the figure
demonstrates there is no clear trend between the two parameters.  The
systems with inclination angle above 30$^{\circ}$ show equivalent
widths that are similar within the 90\% confidence interval with the
mean value of $\sim$50 eV.  In our sample there is only one source
below 30$^{\circ}$ (V603~Aql) that shows the highest equivalent width
of 162$^{+99}_{-65}$ eV.  We need a larger number of sources with
inclinations below 30$^{\circ}$ to explore any relation between the
two parameters.  However, it gives some indication that the systems
with very low inclination (below 20$^{\circ}$) may have a strong
fluorescent line whereas those with an inclination above 70$^{\circ}$
have a weak line.  On the other hand, \cite{hakala04} reported that a
dwarf nova YZ~Cnc has a low inclination angle and still lacks the 6.4
keV fluorescent emission line.  Also \cite{baskill05} found prominent
fluorescent lines in only 4 CVs out of 34 with no correlation of the
line strength and the system inclination.  Thus the reason for the
strength (or weakness) of this line in the non-magnetic CVs remains
unclear.

The old nova V603~Aql shows a strong fluorescent line with the highest
EW of 162$^{+99}_{-65}$ eV among all sources (see \S 3.1).
\cite{mukai05} reports a similar EW for this line and attribute it
to the reflection from a surface that subtends an angle of 2$\pi$ as
seen from the primary X-ray source assuming a solar abundance of Fe.
The line is found to be broadened with a best fit value of line width
$\sigma \sim$37$^{+24}_{-23}$ eV that corresponds to about
1730$^{+1130}_{-1050}$ km s$^{-1}$.  The broadening in this line can
be due to thermal broadening, Doppler broadening or Compton scattering
in the material responsible for reflection.  The natural width of this
line is few eV \citep{george91} and any broadening due to the thermal
motion of the emitting atoms is 0.4(T/$10^{6}$)$^{1/2}$ eV, which is very
small compared to the resolution of the HEG.  Hence the observed
broadening can only be due to either the Doppler effect, Compton
scattering or both.

V426~Oph shows a much weaker fluorescent line emission with an EW of
39$^{+22}_{-18}$ eV when compared with the previously reported value
of 185$\pm$40 eV by \cite{baskill05} using \emph{ASCA} data
taken during 1994 September 18.  However, they found that the
inclusion of a reflection continuum reduced the plasma temperatures
and the EW of the 6.4 keV line.

SS~Cyg, with the highest signal-to-noise ratio, shows a prominent line
at 6.4 keV with EW of 59$^{+25}_{-20}$ eV.  Previous \emph{ASCA} and
\emph{Ginga} data of SS~Cyg show the presence of significant
reflection in their spectra \citep{done97}.  This line is also present
in SS~Cyg during outbursts with a similar EW but their line centers
are red-shifted by 50 eV.  This shift corresponds to a very high
velocity of about 2300 km s$^{-1}$ during the outburst.  An
outflowing wind with such a high velocity has been reported from
SS~Cyg during an outburst \citep{mauche04}.  Also the presence of a
high velocity wind in CVs has been reported previously by
\cite{prinja95} using high resolution IUE observations. We suggest
that the observed high velocity probably arises from the wind that is
moving away from the system during the outburst.

\cite{ezuka99} have shown that apart from reflection from the white dwarf
surface, N$_{H}$ along the line of sight also contributes to the EW of
the fluorescent iron line.  The observed value of N$_{H}$ for these
sources vary in the range of 10$^{20-22}$ cm$^{-2}$
\citep{mukai03,homer04,perna03,pandel05}.  If N$_{H}$ is the only
contributor to the fluorescent Fe line, then for such a low value of
N$_{H}$, the expected value of EW is $\lesssim$10 eV assuming that the
source is surrounded by a uniform absorber at a 4$\pi$ solid angle.
However, the observed best fit value of the EW varies in the range of
23--162 eV (see Table \ref{tbl-2}) indicating the dominance of the
reflection process over the absorption, resulting in the observed
fluorescent Fe line.

\subsection{Quiescence vs. Outburst States}
As mentioned in \S 2, two of the six sources, viz., U~Gem and SS~Cyg,
are observed in both quiescence and outburst.  Such observations
provide an opportunity to study the response of the system with
changes in its intensity.  As shown in Figs. \ref{fig1} \& \ref{fig2},
the three Fe emission lines have considerably different profiles
during the two states.
In particular, the profile of the Fe XXV triplet is significantly different
during the two states.  In U~Gem during quiescence, this line has a
double humped profile reflecting the dominance of the $r$ and $f$
lines over the $i$ components.  During the outburst, the line shows a
broad symmetric Gaussian profile indicating the presence of high
velocity material near the white dwarf. The common line
width ($\sigma$) for $r$, $i$ and $f$ lines is 55$^{+13}_{-19}$ eV,
which corresponds to a velocity of 2460$^{+580}_{-850}$ km s$^{-1}$.

A ``flat-top'' profile is observed for SS~Cyg during an outburst.
Such a profile has been reported for N VII line in O-type
star, $\zeta$~Puppis using \emph{Chandra} HETG data and attributed to small 
velocity gradient at the larger radii in the outflowing winds in the star 
\citep[see][]{cassi01,kahn01}.  
The failure to find a common shift for the $r$, $i$ and $f$ lines 
during the outburst, and the unusual result of a blue-shifted resonance 
component and red-shifted $i$ and $f$ components, is most likely just
symptomatic of the non-gaussian shape of the line profile.
This could be due to velocity-smeared bulk motion of
material present either in a Keplerian orbit around the white dwarf or
a wind flowing away from the system.  \cite{mauche04} reported the
presence of a high velocity wind ($v\approx$2500 km s$^{-1}$) in
SS~Cyg during an outburst observed with the Low Energy Transmission
Grating (LETG) data.  On the other hand, the spectra of SS~Cyg in
quiescence do not show any shift or broadening in the Fe XXV triplet.

Flat top or a double peaked line profiles have also been observed 
in active galactic nuclei, and interpreted as due to
the material flowing away from the central object.  For example, a Seyfert-1
Galaxy Mrk~509 has been observed to show such a line profile from
outflowing gas \citep{phillips83}.  Under favorable viewing
conditions one can see both the material moving away and approaching
towards the observer that can produce such a line profile.  
However, the present data cannot resolve this issue.
It is also possible that some higher-order
effects like optical depth and radiation transfer may be playing a
role in producing the observed line profile.  
In addition, the exposure time of 36 and 60 ksec (see Table~\ref{tbl-1})
of SS~Cyg during 12 and 14 September 2000 covers $\sim$1.5 and 2.5 
orbital cycles with an orbital period of $\sim$6.6 hrs.  The discordant 
wavelength shift is probably caused by the long exposure, during which 
the conditions in the emitting region might have changed significantly.
The summed line may represent a merger of a broad 
range of physical conditions.  Detailed radiative
transfer calculations are required to account for the higher-order effects, 
which is not within the scope of the present paper.

It has been observed during an optical outburst of SS~Cyg that the
hard X-ray flux ($>$3 keV) is suppressed and the spectrum softens,
whereas the X-ray flux is high and the spectrum is harder during the
optical quiescence \citep[][and references therein]{mcgowan04}.  The
observed X-ray flux in SS~Cyg in the HEG band of 0.4--8 keV is 1.8
$\times$ 10$^{-10}$ ergs cm$^{-2}$ s$^{-1}$ during quiescence and it
is 4.0 $\times$ 10$^{-11}$ and 4.8 $\times$ 10$^{-11}$ ergs cm$^{-2}$
s$^{-1}$ during the outburst of 12 and 14 September 2000,
respectively.  This suggests that for SS~Cyg during an optical
outburst the X-ray flux is suppressed by a factor of $\sim$3--4
supporting the previous observations.  On the other hand U~Gem shows
the opposite behavior, when it goes in to optical outburst its X-ray
flux in 0.4--8 keV range increases by a factor of 2--3.

\subsection{Comparison with Magnetic CVs}
Prominent Fe K$_{\alpha}$ emission lines have been observed in
magnetic CVs (MCVs).  The Fe XXV and Fe XXVI lines are believed to
originate in a hot plasma in the post-shock region and the appearance
of the fluorescent Fe line is attributed to reflection from the white
dwarf surface or the cool pre-shock material \citep{ezuka99}.  A study
of Fe K${\alpha}$ complex in magnetic CVs has been reported by
\cite{hellier04} using the \emph{Chandra} HETG data.  They use a power
law and three Gaussian components to model the Fe K$_{\alpha}$
emission lines in a sample of five MCVs.  They find that the Fe
XXVI line is red-shifted by 260 km s$^{-1}$ from a simultaneous fit to
the spectra of five MCVs and suggested that the presence of the
satellite lines can cause the small observed red-shift.  For
non-magnetic CVs, we have not observed any detectable shift in the
line energy of the Fe XXVI line during both quiescence and outburst
suggesting an absence of any Doppler shift in
these systems.  \cite{hellier04} find that the Fe XXVI line is, in
general, broadened by 1000 km s$^{-1}$ in MCVs.  For non-magnetic CVs,
only U~Gem shows a significant broadening in Fe XXVI line by
1400$^{+840}_{-600}$ km s$^{-1}$ in quiescence.

\cite{hellier04} also reported that the two intermediate polars
(IPs), AO~Psc and EX~Hya, have a much stronger resonance line of the Fe
XXV triplet than is observed in the polar AM~Her.  In our study, this
line is also found to be strong like that observed in the two IPs in which the
accretion takes place via a partial or truncated disk.  The common
thread between the IPs and DNe is an accretion disk that is absent in
polars.  This might provide some clue for the dependence of the Fe XXV
resonance line strength on the presence of accretion disk or the
strength of magnetic field.  Observations of a large number of
CVs with a high resolution X-ray instrument can help to investigate this
further.

\cite{ezuka99} find that the reflection of hard X-rays from the white
dwarf generally makes a significant contribution to the fluorescent
iron line in MCVs having a weak line-of-sight absorber
(N$_{H}<$10$^{22}$ cm$^{-2}$).  This suggests that in these systems
the production mechanism of the fluorescent Fe line is independent of
the magnetic nature of the system.  \cite{hellier04} have detected a
red wing of the fluorescent Fe line extending to 6.33 keV in the IP GK~Per.
This broadening corresponds to a Doppler shift of up to 3700
km s$^{-1}$.  They suggest that this arises from the pre-shock
material that is free-falling onto the white dwarf.  We have detected
a symmetrically broadened fluorescent Fe line in V603~Aql with a
broadening of $\sim$1700 km s$^{-1}$, which is probably due to Doppler
broadening or Compton scattering of material in the accretion disk.
The red-shift of up to 2300 km s$^{-1}$ detected in SS~Cyg during
outburst is attributed to the wind flowing away from the system unlike
the free-falling material in MCVs.

\section{Summary}
\begin{enumerate}

\item The presence of a strong Fe XXV triplet is a common feature in
the hard X-ray spectra of non-magnetic CVs irrespective of their
intensity state.  During the quiescent state the $r$ component is
stronger than the $i$ and $f$ components indicating a temperature of
$>$3 $\times 10^{7}$ K for emitting plasma.

\item The G-ratio is close to unity for all sources indicating the
presence of collisionally ionized plasma in these sources during
quiescence.  The measured values of R-ratio for Fe~XXV from the present
HEG data do not allow us to constrain the plasma densities in the sources
studied here.

\item During outburst the $r$, $i$ and $f$ components of the He-like line in
U~Gem are found to be significantly broadened with a velocity of 
2460$^{+580}_{-850}$ km s$^{-1}$.  The presence of high velocity material
($\approx$1200 km s$^{-1}$) in the form of a Keplerian disk or wind
flowing away from the system is indicated for SS~Cyg during an
outburst confirming the previous report by \cite{mauche04} using the
\emph{Chandra} LETG data.

\item For SS~Cyg the Fe XXVI/Fe XXV line ratio provides a higher
ionization temperature during quiescence than in the outburst state.
This supports other observations that show the X-ray spectrum of
SS~Cyg to be softer in outburst than quiescence \citep[][and
references therein]{mcgowan04}.

\item A study of Fe XXV resonance line in a large sample of CVs
including polars, IPs and DNe may provide some clue as to the
dependence of the resonance line strength on the presence of an
accretion disk or the magnetic nature of the system.

\item The presence of a prominent fluorescent iron line in V603~Aql,
V426~Oph and SS~Cyg indicates the presence of significant reflection
in these systems during quiescence.  V603~Aql shows a significantly 
broadened fluorescent Fe line corresponding to a velocity of 
1730$^{+1130}_{-1050}$ km s$^{-1}$. For SS~Cyg, the fluorescent Fe 
line is red-shifted with energies of 6.35$\pm$0.02 and 6.35$\pm$0.01 keV,
during the two outburst observations done on 2000 September 12 and 14 
respectively, indicating a velocity of about 2300$^{+980}_{-900}$  
and 2300$^{+500}_{-440}$ km s$^{-1}$ for fluorescent material.

\item The EW of the fluorescent Fe line during quiescence in all
sources is consistent with reflection from the white dwarf surface
that subtends an angle of about 2$\pi$ at the X-ray source.

\end{enumerate}

\acknowledgments
This research has made use of data obtained from the Chandra Data
Archive (CDA) which is a part of the Chandra X-Ray Observatory Science
Center (CXC) and is operated for NASA by the Smithsonian Astrophysical
Observatory.  Authors wish to thank the anonymous referee for the
valuable suggestions that helped in improving the paper. V.R.R is
pleased to acknowledge partial support from the Kanwal Rekhi
Scholarship of the TIFR Endowment Fund.  The research of E.M.S is
supported by NASA contract number NAS8-39073 to SAO to operate the
Chandra X-ray Observatory.  The research of P.E.B. is supported by
NASA contract NNG04GL18G.



\clearpage

\clearpage

\begin{deluxetable}{cccll}
\tabletypesize{\scriptsize}
\tablecaption{\textit{Chandra} HETG observation log for non-magnetic CVs. \label{tbl-1}}
\tablewidth{0pt}
\tablehead{
\colhead{source} & \colhead{Obs-id} & \colhead{Exposure} & \colhead{Date} & \colhead{State} \\
                 &                  & \colhead{(ksec)}   &                &
}
\startdata
SS~Cyg    & 646  & 48  & 2000 August 24    & Quiescence    \\
          & 2307 & 36  & 2000 September 12 & Outburst     \\
          & 648  & 60  & 2000 September 14 & Outburst    \\
U~Gem     & 647  & 96  & 2000 November 29  & Quiescence  \\
          & 3767 & 62  & 2002 December 26  & Outburst    \\
V603~Aql  & 1901 & 64  & 2001 April 19     & Quiescence  \\
V426~Oph  & 2671 & 45  & 2002 May 30       & Quiescence   \\
WX~Hyi    & 2670 & 50  & 2002 July 28      & Quiescence  \\
          & 3721 & 50  & 2002 July 25      & Quiescence   \\
SU~UMa    & 2680 & 49  & 2002 April 13     & Quiescence  \\
          & 3478 & 23  & 2002 April 12     & Quiescence  \\
\enddata




\end{deluxetable}


\clearpage

\begin{deluxetable}{ccccccccccc}
\tablecolumns{11}
\tabletypesize{\scriptsize}
\tablecaption{Fe K${\alpha}$ line measurements for non-magnetic CVs in quiescent and outburst states. \label{tbl-2}}
\tablewidth{0pt}
\tablehead{
\colhead{source} & \multicolumn{5}{c}{Equivalent width (eV)} &
\multicolumn{5}{c}{Line Flux ($\times 10^{-5}$ ph cm$^{-2}$ s$^{-1}$)}  \\
\cline{2-6} \cline{7-11}
\colhead{} & \colhead{fluorescent} & \multicolumn{3}{c}{He-like} &
\colhead{H-like} &
\colhead{fluorescent} & \multicolumn{3}{c}{He-like} & \colhead{H-like}  \\
\colhead{} & \colhead{} & \colhead{r} & \colhead{i} &
\colhead{f} & \colhead{} & \colhead{} & \colhead{r} & \colhead{i} & \colhead{f}
}
\startdata
\cutinhead{Quiescent State}
SS~Cyg   & 59$^{+25}_{-20}$ & 35$^{+20}_{-18}$ & 22$^{+20}_{-16}$ &
20$^{+12}_{-10}$ & 65$^{+20}_{-17}$ & 11.1$^{+2.6}_{-2.4}$ &
6.7$^{+4.0}_{-3.3}$ & 4.5$^{+3.9}_{-3.3}$ & 4.2$^{+2.5}_{-2.1}$ &
10.9$^{+3.2}_{-2.9}$  \\
U~Gem    & 46$^{+31}_{-24}$ & 134$^{+77}_{-61}$ & 38$^{+65}_{-38}$ &
124$^{+50}_{-40}$ & 181$^{+71}_{-58}$ & 0.7$^{+0.5}_{-0.4}$ &
2.2$^{+1.2}_{-1.0}$ & 0.7$^{+1.2}_{-0.7}$ & 2.0$^{+0.8}_{-0.6}$ &
2.6$^{+1.0}_{-0.8}$ \\
V603~Aql & 162$^{+99}_{-65}$ & 98$^{+80}_{-73}$ & 47$^{+74}_{-47}$ &
35$^{+40}_{-30}$ & 69$^{+65}_{-49}$ & 2.5$^{+1.1}_{-1.5}$ &
1.6$^{+1.3}_{-1.2}$ & 0.8$^{+1.3}_{-0.8}$ & 0.7$^{+0.7}_{-0.6}$ &
0.9$^{+0.9}_{-0.6}$ \\
V426~Oph & 39$^{+22}_{-18}$ & 90$^{+47}_{-39}$ & 59$^{+45}_{-37}$ &
48$^{+26}_{-21}$ & 136$^{+47}_{-39}$ & 2.6$^{+1.5}_{-1.2}$ &
6.2$^{+3.3}_{-2.7}$ & 4.4$^{+3.3}_{-2.7}$ & 3.6$^{+1.9}_{-1.5}$ &
8.3$^{+2.9}_{-2.3}$  \\
WX~Hyi   & 38$^{+50}_{-30}$ & 115$^{+100}_{-84}$ & 103$^{+122}_{-91}$ &
93$^{+74}_{-50}$ & 74$^{+50}_{-45}$ & 0.5$^{+0.6}_{-0.4}$ &
1.7$^{+1.6}_{-1.2}$ & 1.5$^{+1.8}_{-1.3}$ & 1.4$^{+1.1}_{-0.8}$ &
1.0$^{+1.0}_{-0.7}$ \\
WX~Hyi   & 55$^{+44}_{-30}$ & 285$^{+110}_{-93}$ & 18$^{+70}_{-18}$ &
81$^{+62}_{-43}$ & 71$^{+60}_{-45}$ & 1.0$^{+0.7}_{-0.5}$ &
3.8$^{+1.6}_{-1.2}$ & 0.3$^{+1.4}_{-0.3}$ & 1.4$^{+1.1}_{-0.8}$ &
1.1$^{+1.0}_{-0.7}$ \\
SU~UMa   & 23$^{+46}_{-23}$ & 139$^{+120}_{-88}$ & 201$^{+145}_{-119}$
& 24$^{+70}_{-24}$ & 0$^{+52}$ & 0.2$^{+0.4}_{-0.2}$ & 0.9$^{+0.8}_{-0.6}$ &
1.2$^{+0.9}_{-0.7}$ & 0.2$^{+0.6}_{-0.2}$ & $<$0.4  \\
\cutinhead{Outburst State}
SS~Cyg & 60$^{+14}_{-14}$ & 70$^{+25}_{-10}$ & 72$^{+19}_{-17}$ &
82$^{+19}_{-19}$ & 27$^{+12}_{-12}$ & 3.8$^{+1.0}_{-1.0}$ &
4.2$^{+1.6}_{-0.5}$ & 4.7$^{+1.4}_{-1.0}$ & 5.3$^{+1.2}_{-1.2}$ &
1.5$^{+0.6}_{-0.6}$     \\
SS~Cyg & 68$^{+26}_{-40}$ & 99$^{+54}_{-36}$ & 60$^{+34}_{-34}$ &
65$^{+40}_{-28}$ & 27$^{+37}_{-24}$ & 3.0$^{+1.2}_{-1.8}$ &
4.2$^{+2.3}_{-1.5}$ & 3.0$^{+1.7}_{-1.7}$ & 3.1$^{+1.9}_{-1.3}$ &
1.2$^{+1.6}_{-1.1}$  \\
U~Gem  & 22$^{+16}_{-14}$ & 26$^{+48}_{-26}$ & 325$^{+117}_{-325}$ &
28$^{+56}_{-28}$ & 65$^{+26}_{-23}$ & 1.5$^{+1.1}_{-0.9}$ &
1.9$^{+3.6}_{-1.9}$ & 17.0$^{+6}_{-17}$ & 2.1$^{+4.2}_{-2.1}$ &
3.9$^{+1.6}_{-1.4}$  \\
\enddata
\tablecomments{The errors are with 90\% confidence interval for a single
parameter ($\Delta\chi^{2}$=2.706).}

\end{deluxetable}

\clearpage

\begin{deluxetable}{cccc}
\tablecolumns{4}
\tabletypesize{\scriptsize}
\tablecaption{Line intensity ratios for Fe XXV triplets and Fe XXVI ions. \label{tbl-3}}
\tablewidth{0pt}
\tablehead{
\colhead{source} & \colhead{G=(f+i)/r} & \colhead{R=f/i} &
\colhead{Fe XXVI/XXV}
}
\startdata
\cutinhead{Quiescent State}
SS~Cyg   & 1.30$^{+1.04}_{-0.87}$ & 0.93$^{+0.98}_{-0.83}$ &
0.71$^{+0.35}_{-0.30}$   \\
U~Gem    & 1.23$^{+0.94}_{-0.70}$ & 2.86$^{+5.03}_{-2.86}$ &
0.53$^{+0.29}_{-0.22}$   \\
V603~Aql & 0.94$^{+1.20}_{-0.94}$ & 0.88$^{+1.67}_{-1.15}$ &
0.29$^{+0.34}_{-0.24}$   \\
V426~Oph & 1.29$^{+0.92}_{-0.75}$ & 0.82$^{+0.75}_{-0.61}$ &
0.58$^{+0.29}_{-0.23}$   \\
WX~Hyi   & 1.71$^{+2.03}_{-1.50}$ & 0.93$^{+1.34}_{-0.97}$ &
0.22$^{+0.25}_{-0.18}$   \\
WX~Hyi   & 0.45$^{+0.51}_{-0.27}$ & 4.67$^{+22.1}_{-5.37}$ &
0.20$^{+0.20}_{-0.14}$   \\
SU~UMa   & 1.55$^{+1.83}_{-1.32}$ & 0.17$^{+0.52}_{-0.19}$ & 
--  \\
\cutinhead{Outburst State}
SS~Cyg   & 2.40$^{+1.01}_{-0.47}$ & 1.13$^{+0.42}_{-0.35}$ &
0.11$^{+0.04}_{-0.05}$   \\
SS~Cyg   & 1.45$^{+1.00}_{-0.73}$ & 1.03$^{+0.86}_{-0.73}$ &
0.12$^{+0.16}_{-0.11}$   \\
U~Gem    & 10$^{+19}_{-10}$ & 0.12$^{+0.25}_{-0.17}$ &
0.19$^{+0.11}_{-0.17}$   \\
\enddata
\tablecomments{The errors are obtained by propagating the 90\%
confidence errors on line fluxes from Table~\ref{tbl-2}.}

\end{deluxetable}


\begin{thebibliography}{}
\bibitem[Arnaud(1996)]{arnaud96} Arnaud, K. A. 1996, in ASP
Conf. Ser. 101, Astronomical Data Analysis Software and Systems V,
ed. G. Jacoby \& J. Barnes (San Francisco: ASP), 17
\bibitem[Baskill, Wheatley, \& Osborne(2005)]{baskill05} Baskill,
D. S., Wheatley, P. J., \& Osborne, J. P. 2005, \mnras, 357, 626
\bibitem[Bautista \& Kallman(2000)]{bau00} Bautista, M. A. \& 
Kallman, T. R. 2000, \apj, 544, 581
\bibitem[Cassinelli et al.(2001)]{cassi01} Cassinelli, J. P., Miller,
M. A., Waldron, W. L., MacFarlane, J. J., \& Cohen, D. H.  2001, \apj,
554, L55
\bibitem[Done \& Osborne(1997)]{done97} Done, C., \& Osborne, J. P.
1997, \mnras, 288, 649
\bibitem[Ezuka \& Ishida(1999)]{ezuka99} Ezuka, H., \& Ishida,
M. 1999, \apjs, 120, 277
\bibitem[Garmire et al.(2003)]{garmire03} Garmire, G. P., Bautz, M. W.,
Ford, P. G., Nousek, J. A., \& Ricker, G. R., Jr. 2003, SPIE, 4851, 28
\bibitem[George \& Fabian(1991)]{george91} George, I. M., \& Fabian,
A. C. 1991, \mnras, 249, 352
\bibitem[Hakala et al.(2004)]{hakala04} Hakala, P., Ramsay, G.,
Wheatley, P., Harlaftis, E. T., Papadimitriou, C. 2004, \aap, 420, 273
\bibitem[Hellier \& Mukai(2004)]{hellier04} Hellier, C., \& Mukai,
K. 2004, \mnras, 352, 1037
\bibitem[Homer et al.(2004)]{homer04} Homer, L., Szkody, P., Raymond,
J. C., Fried, R. E., Hoard, D. W., Hawley, S. L., Wolfe, M. A.,
Tramposch, J. N., \& Yirak, K. T. 2004, \apj, 610, 991
\bibitem[Kahn et al.(2001)]{kahn01} Kahn, S. M., et al.	2001, \aap,
365, L312
\bibitem[Lasota et al.(2001)]{lasota01} Lasota, J. P. 2001, New
Astron. Rev., 45, 449
\bibitem[Markert et al.(1994)]{markert94} Markert, T. H., Canizares,
C. R., Dewey, D., McGuirk, M., Pak, C. S., \& Schattenburg,
M. L. 1994, Proc. SPIE, 2280, 168
\bibitem[Mauche(2004)]{mauche04} Mauche, C. 2004, \apj, 610, 422
\bibitem[Mauche et al.(2005)]{mauche05} Mauche, C. W., Wheatley, P. J.,
Long, K. S., Raymond, J. C., \& Szkody, P. 2005, ASP Conf. series,
vol. 330, 355, Eds: J. M. Hameury \& J. P. Lasota
\bibitem[McGowan, Priedhorsky, \& Trudolyubov(2004)]{mcgowan04}
McGowan, K. E., Priedhorsky, W. C., \& Trudolyubov, S. P. 2004,
\apj, 601, 1100
\bibitem[Mewe et al.(1985)]{mewe85} Mewe, R., Gronenschild,
E. H. B. M., \& van den Oord, G. H. J. 1985, \aaps, 62, 197
\bibitem[Mukai \& Shiokawa(1993)]{mukai93} Mukai, K., \& Shiokawa,
K. 1993, \apj, 418, 863
\bibitem[Mukai et al.(2003)]{mukai03} Mukai, K., Kinkhabwala, A.,
Peterson, J. R., Kahn, S. M., \& Paerels, F. 2003, \apjl, 586, L77
\bibitem[Mukai \& Orio(2005)]{mukai05} Mukai, K., \& Orio, M. 2005,
\apj, 622, 602
\bibitem[Nousek \& Shue(1989)]{nousek89} Nousek, J. A., \& Shue,
D. R. 1989, \apj, 342, 1207
\bibitem[Oelgoetz \& Pradhan(2001)]{oelgoetz01} Oelgoetz, J., \&
Pradhan, A. K. 2001, \mnras, 327, L42
\bibitem[Pandel et al.(2005)]{pandel05} Pandel, D., Cordova, F. A.,
Mason, K. O., \& Priedhorsky, W. C. 2005, \apj, 626, 396
\bibitem[Perna et al.(2003)]{perna03} Perna, R., McDowell, J., Menou,
K., Raymond, J., \& Medvedev, M. V. 2003, \apj, 598, 545
\bibitem[Phillips et al.(1983)]{phillips83} Phillips, M. M., Baldwin,
J. A., Atwood, B., \& Carswell, R. F. 1983, \apj, 274, 558
\bibitem[Prinja \& Rosen(1995)]{prinja95} Prinja, R. K., \& Rosen,
S. R.	1995, \mnras, 273, 461
\bibitem[Ramsay et al.(2001)]{ramsay01} Ramsay, G., Cordova, F.,
Cottam, J., Mason, K., Osborne, J., Pandel, D., Poole, T., \&
Wheatley, P.  2001, \aap, 365, L294
\bibitem[Szkody et al.(2002)]{szkody02} Szkody, P., Nishikida, K.,
Raymond, J. C., Seth, A., Hoard, D. W., Long, K., \& Sion, E. M. 2002,
\apj, 574, 942
\bibitem[Warner(1995)]{war95} Warner, B. 1995, Cataclysmic Variables
(Cambridge: Cambridge University Press)
\bibitem[Wheatley, Mauche, \& Mattei(2003)]{wheatley03} Wheatley, P.
J., Mauche, C. W., and Mattei, J. A. 2003, \mnras, 345, 49
\bibitem[Wiesskopf et al.(2002)]{weiss02} Wiesskopf, M. C., Brinkman,
B., Canizares, C., Garmire, G., Murray, S., \& Van Speybroeck,
L. P. 2002, \pasp, 114, 1

\end{thebibliography}
\end{document}